\documentclass{emulateapj}

\slugcomment{Submitted to {\it The Astrophysical Journal}}
\shorttitle{Follow-up Observations of Two Pulsating Hot DQ Stars} 
\shortauthors{P. Dufour et al.}

\newcommand{\gta}{\lower 0.5ex\hbox{$ \buildrel>\over\sim\ $}}
\newcommand{\lta}{\lower 0.5ex\hbox{$ \buildrel<\over\sim\ $}}

\begin{document}

\title{Follow-up Observations of the Second and Third Known Pulsating
  Hot DQ White Dwarfs}

\author{P. Dufour\altaffilmark{1,2}, E.M. Green\altaffilmark{1}, G. Fontaine\altaffilmark{2}, P. Brassard\altaffilmark{2}, M. Francoeur\altaffilmark{2} and M. Latour\altaffilmark{2}}

\altaffiltext{1}{Steward Observatory, University of Arizona, 933 North
  Cherry Avenue, Tucson, AZ 85721; bgreen@as.arizona.edu}
\altaffiltext{2}{D\'epartement de Physique, Universit\'e
    de Montr\'eal, Montr\'eal, QC H3C 3J7, Canada; dufourpa@astro.umontreal.ca, fontaine@astro.umontreal.ca, brassard@astro.umontreal.ca, myriam@astro.umontreal.ca, marilyn@astro.umontreal.ca}

\begin{abstract}

We present follow-up time-series photometric observations that confirm
and extend the results of the significant discovery made by Barlow et
al.\ (2008) that the Hot DQ white dwarfs SDSS J220029.08$-$074121.5 and 
SDSS J234843.30$-$094245.3 are luminosity variable. These are the second
and third known members of a new class of pulsating white dwarfs, after
the prototype SDSS J142625.71+575218.3 (Montgomery et al.\ 2008). We find
that the light curve of SDSS J220029.08$-$074121.5 is dominated by an 
oscillation at 654.397$\pm$0.056~s, and that the light pulse folded on
that period is highly nonlinear due to the presence of the first and
second harmonic of the main pulsation. We also present evidence for the
possible detection of two additional pulsation modes with low amplitudes
and periods of 577.576$\pm$0.226~s and 254.732$\pm$0.048~s in that
star. Likewise, we find that the light curve of SDSS J234843.30$-$094245.3 
is dominated by a pulsation with a period of 1044.168$\pm$0.012~s, but
with no sign of harmonic components. A new oscillation, with a low
amplitude and a period of 416.919$\pm$0.004~s, is also probably detected
in that second star. We argue, on the basis of the very different folded
pulse shapes, that SDSS J220029.08$-$074121.5 is likely magnetic, while
SDSS J234843.30$-$094245.3 is probably not.  

\end{abstract}


\section{INTRODUCTION}

The existence of a new class of luminosity-variable white dwarfs
was confirmed recently by the very important contribution of Barlow et
al.\ (2008) who reported on their discovery of periodic variations in the
light curves of two carbon-atmosphere (Hot DQ) stars from the Sloan
Digital Sky Survey. This followed the initial discovery of Montgomery et
al.\ (2008) that the Hot DQ white dwarf SDSS J142625.71+575218.3
(hereafter SDSS J1426+5752) shows luminosity variations dominated by a
periodicity at 417.7~s along with its first harmonic. That harmonic proved 
remarkable in that it reached the unusually high amplitude of some 40\% of
that of the main oscillation. During an engineering run, Barlow et
al.\ (2008) used the Goodman Spectrograph in imaging mode at the 4.1~m
SOAR Telescope to monitor in broadband the light curves of two Hot DQ
stars not previously investigated by Montgomery et al.\ (2008). The
latter had found one variable star out of a sample of six objects of this
kind. On the basis of 5.80~h of observations, Barlow et al.\ (2008) found
that SDSS J220029.08$-$074121.5 (hereafter SDSS J2200$-$0741) shows
strong similarities in its photometric properties with SDSS J1426+5752
in that it displays a light curve dominated by a main oscillation plus
its first harmonic. The main oscillation in SDSS J2200$-$0741 was
determined to have a period of about 656~s, and the amplitude of its 
first harmonic, even more unusually, is nearly comparable to that of the 
dominant periodicity. Barlow et al.\ (2008) also
found, on the basis of a further 4.71~h of observations, a significant
variation with a period of about 1052~s in the light curve of SDSS
J234843.30$-$094245.3 (hereafter SDSS J2348$-$0943), but with no sign of
a strong harmonic or other periodicities within their detection limit. 

The results of Barlow et al.\ (2008), combined with that of Montgomery et
al.\ (2008), indicate that three Hot DQ white dwarfs, out of a total of
eight investigated so far, show periodic luminosity variations. These 
targets all come from the very small family of nine Hot DQ stars 
found previously by Dufour et al.\ (2007, 2008a) in a sample of
nearly 10,000 spectroscopically identified white dwarfs. The Hot DQ
stars form a newly-discovered and exceedingly rare type of white
dwarfs. The preliminary atmospheric analysis carried out by Dufour et
al.\ (2008b) indicates that all the Hot DQ white dwarfs fall in a narrow
range of effective temperature, between about 18,000~K and 24,000~K, and
that they have atmospheric carbon-to-helium number ratios ranging from 1
to upward of 100. Depending on the exact values of the effective
temperature, surface gravity, and envelope/atmosphere chemical
composition, Fontaine, Brassard, \& Dufour (2008) demonstrated that
low-order and low-degree gravity-mode oscillations could be excited in
models of Hot DQ stars in a way very similar to what is encountered in
pulsating white dwarfs of the V777 Her and ZZ Ceti types. Pulsational
instabilities thus provide a most natural explanation for the luminosity
variations seen in some of the Hot DQ stars (and see Fontaine et
al.\ 2009 for more details).  

Montgomery et al.\ (2008) also proposed pulsational instabilities as
their preferred explanation for the luminosity variations observed in
SDSS J1426+5752, but they alternatively suggested that these variations 
could be associated with photometric activity in a carbon 
analog of AM CVn, the prototype of the helium-transferring cataclysmic
variables (see, e.g., Warner 1995 on the properties of AM CVn systems). 
This was based on the observation that the light curve of
SDSS J1426+5752 folded on the dominant periodicity of 417.7~s resembles
that of AM CVn itself, exhibiting a flatter maximum than minimum,
which is the opposite of what is seen in large amplitude pulsating white
dwarfs. The first harmonic in a large amplitude pulsating white dwarf is
quite generally nearly in phase with the main mode at light maximum, while
it tends to be in antiphase in SDSS J1426+5752 and in AM CVn systems,
thus producing the different pulse shapes. We note that Barlow et
al.\ (2008) invoked this pulse shape argument again in their paper,
arguing that pulsational instabilities are not necessarily to be preferred
over the interacting binary hypothesis as an explanation for the
luminosity variations seen in the variable Hot DQ stars. However, in the
meantime, Green, Dufour, \& Fontaine (2009) have presented a detailed
analysis based on follow-up photometric and spectroscopic observations
of SDSS J1426+5752. At the end of their exercise, they unequivocally
ruled out the interacting binary hypothesis and concluded instead that,
indeed, the luminosity variations seen in SDSS J1426+5752 are caused by
$g$-mode pulsations as in other pulsating white dwarfs. Since their
results can readily be generalized to other variable Hot DQ stars, 
from now on we consider variability in these stars to be caused by
pulsational instabilities.

In this paper, we present the results of follow-up CCD photometry in
integrated light gathered at the Steward Observatory 1.55~m Kuiper
Telescope for both SDSS J2200$-$0741 and SDSS J2348$-$0943. Given the
importance of the findings of Barlow et al.\ (2008) in the establishment
of a new class of pulsating stars, we felt that rapid confirmation of
their results should be given high priority. We thus carried out our
observations as soon as we could after learning of the Barlow et al.\
(2008) results through collaborators. Despite the relative faintness of
the target stars, $g$ = 17.70 for SDSS J2200$-$0741 and $g$ = 19.00 for
SDSS J2348$-$0943, we were able to collect quite useful data. We present
our analysis of these data in what follows. 

\section{SDSS J2200$-$0741}

\subsection{Observations}

We followed the same procedure as that described in Green et al.\ (2009)
for their follow-up photometric observations of SDSS J1426+5752. Hence,
we used the Kuiper/Mont4K combination\footnote{Please consult the
  following web site: 
  http://james.as.arizona.edu/$\sim$psmith/61inch/instruments.html for more
  details, if interested.} 
to gather some 13.29~h of time-series observations over a baseline of
52.37~h (three consecutive nights). The observations were taken on dark
nights through a broadband Schott 8612 filter, and an effective exposure
time of 65~s (on average) was used as a compromise between the S/N and
the need to adequately sample the luminosity variations reported by
Barlow et al.\ (2008). The data set we obtained for SDSS J2200$-$0741 is
characterized by a formal temporal resolution of 5.30 $\mu$Hz and a
duty cycle of 25.4\%. Details of the observations are provided in Table~1.

% Table 1
\begin{deluxetable}{cccc}
\tablewidth{0pt}
\tablecaption{Journal of Observations for SDSS J2200$-$0741}
\tablehead{
\colhead{Date} &
\colhead{Start of Run} &
\colhead{Number of Frames} &
\colhead{Length}\\
\colhead{(UT)} &
\colhead{(HJD2454710+)} &
\colhead{} &
\colhead{(s)}
}
\startdata
2008 Sep 03 & 2.7018786 & 249 & 16,089\\
2008 Sep 04 & 3.7084887 & 240 & 15,433\\
2008 Sep 05 & 4.6947030 & 254 & 16,337\\
\enddata
\end{deluxetable}

Figure 1 shows the three nightly light curves that were obtained. The
original images were reduced using standard IRAF photometric reduction
tasks.  We set the photometric aperture size individually for each 
frame to 2.0 times the FWHM in that image, and computed differential 
light curves of SDSS J2200$-$0741 relative to the average of 
three suitable comparison stars well distributed around the target. 
Final detrending of the effects of differential extinction was made
through spline fitting. A zoomed-in view of the light curve gathered on
September 4 (2008) is provided in Figure 2. Given the magnitude of the
target ($g$ = 17.70) and the small aperture of the Kuiper Telescope, the
overall quality of the light curves is gratifying. We attribute this to
the excellent sensitivity of the CCD, the relatively symmetrical distribution 
of the reference stars on the sky, and our optimized data pipeline. 
Without even the benefit of folding, Figures 1 and 2 clearly reveal that
the light curve of SDSS J2200$-$0741 has flatter maxima than minima. 
This is similar to the behavior of SDSS J1426+5752, as already pointed out
by Barlow et al.\ (2008).

\begin{figure}[!ht]
\plotone{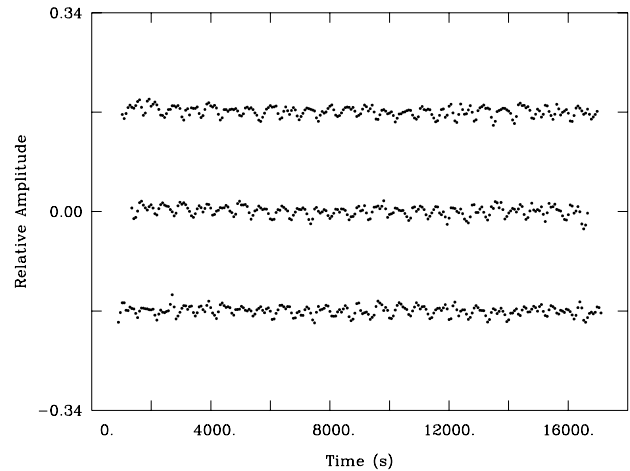}
\caption{All light curves obtained for SDSS J2200$-$0741
using the Mont4K CCD camera mounted on the Steward Observatory 1.55~m 
Kuiper telescope. The data have been shifted arbitrarily along the x and
y axes for visualization purposes. They are expressed in units of
fractional brightness intensity and seconds. From top to bottom, the
curves refer to the nights of UT September 3, September 4, and September
5 (2008). For details, see Table 1.}
\end{figure}

\begin{figure}[!ht]
\plotone{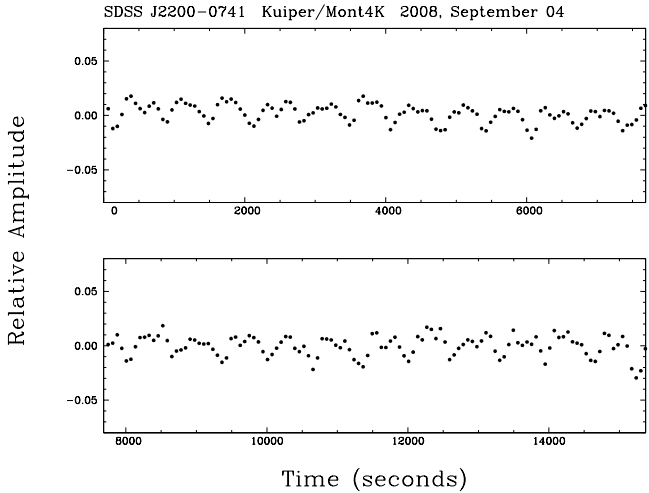}
\caption{Expanded view of the light curve of SDSS
J2200$-$0741 obtained on September 4, 2008. The units are the same as
those used in Fig.\ 1. One can clearly see that the minima are generally
sharper than the maxima.}
\end{figure}

\subsection{Frequency Analysis}

The time-series photometry gathered for SDSS J2200$-$0741 was analyzed in a
standard way using a combination of Fourier analysis, least-squares fits
to the light curve, and prewhitening techniques (see, e.g., Bill\`eres
et al.\ 2000 for more details). Figure 3 shows the Fourier amplitude
spectrum of the full data set in the 0$-$7.5 mHz bandpass (upper curve)
and the resulting transforms after prewhitening of two (middle curve)
and four frequency peaks (lower curve). Our results reveal that the
light curve of SDSS J2200$-$0741 is dominated by an oscillation with a
period of around 654.4~s along with its first harmonic at 327.2~s. The
harmonic has an amplitude of nearly 82\% of that of the main oscillation, 
which readily explains the highly nonlinear shape of the light pulses
seen in Figures 1 and 2. This is in excellent agreement with, and thus
provides a confirmation of, the findings of Barlow et al.\ (2008). In
addition, our higher S/N data also suggest the possible presence  
of two additional oscillations with low amplitudes and periods of around
577.6~s and 254.7~s, as can be seen in Figure 3. These are not
harmonically related to the main pulsation. It is also quite likely
that there are additional frequency components in the light curve of
SDSS J2200$-$0741, but we could not detect them due to the finite
sensitivity of our observations.

\begin{figure}[!ht]
\plotone{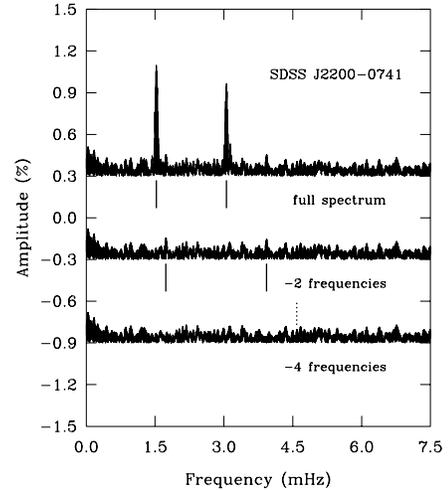}
\caption{Fourier transform of the entire data set in the
0$-$7.5 mHz range (upper curve). The lower transforms show the
successive steps of prewhitening by the two strongest frequencies (the
654.397~s peak and its first harmonic), and finally by all four
frequencies that we isolated. The dotted vertical line segment
associated with the lower curve indicates the location of the second
harmonic of the main periodicity.}
\end{figure}

The results of our frequency analysis are summarized in Table 2, where we
list the basic characteristics of the four oscillations that we isolated.
In addition, Table 2 gives the derived parameters of the second harmonic
(218.1~s) of the main periodicity, an oscillation that we uncovered
after the fact (see below). Note that the phase given in the table is
relative to an arbitrary point in time; in our case, the beginning of
the first run on UT 3 September 2008. The uncertainties on the period,
frequency, amplitude, and phase of each oscillation listed in the
table were estimated with the method put forward by Montgomery \&
O'Donoghue (1999). We point out, in this context, that the uncertainties
on the amplitudes and phases obtained by our least-squares fits during the
prewhitening stage were identically the same as those derived with the
Montgomery \& O'Donoghue (1999) method. A basic quantity in the latter
approach is the average noise level in the bandpass of interest.  When we
computed this mean value from the residual Fourier transform (the
lower curve in Figure 3), the mean noise level in the 0$-$7.5 mHz interval 
turned out to be 0.045\% of the mean brightness of the star. 

% Table 2
\begin{deluxetable}{cccc}
\tablewidth{0pt}
\tablecaption{Harmonic Oscillations Detected in the Light Curve of SDSS
  J2200$-$0741 }
\tablehead{
\colhead{Period} &
\colhead{Frequency} &
\colhead{Amplitude} &
\colhead{Phase}\\
\colhead{(s)} &
\colhead{(mHz)} &
\colhead{(\%)} &
\colhead{(s)}
}
\startdata
654.397$\pm$0.056 & 1.5281$\pm$0.0001 & 0.800$\pm$0.036 & 418.7$\pm$4.8\\
327.218$\pm$0.017 & 3.0560$\pm$0.0002 & 0.655$\pm$0.036 & 254.3$\pm$2.9\\
218.097$\pm$0.052 & 4.5851$\pm$0.0011 & 0.096$\pm$0.036 & 201.0$\pm$13.2\\
577.576$\pm$0.226 & 1.7314$\pm$0.0007 & 0.155$\pm$0.036 & 134.7$\pm$21.7\\
254.732$\pm$0.048 & 3.9257$\pm$0.0007 & 0.145$\pm$0.036 & 135.9$\pm$10.2\\
\enddata
\end{deluxetable}

Given this noise value, and following standard procedure, our
possible detection of the low amplitude 577.6~s periodicity can be seen
as a 3.4 $\sigma$ (0.155/0.045) result. An evaluation of the false alarm
probability following the method of Kepler (1993) indicates that there
is a $\sim$9.9\% chance that this frequency peak is due to noise, a
nonnegligible value. Likewise, our possible detection of the 
254.7~s periodicity is a 3.2~$\sigma$ (0.145/0.045) result. In that
case, the false alarm probability rises to $\sim$27.1\%, making the
detection even more insecure. However, we note that, outside the very low
frequency regime, only these two peaks have amplitudes larger than 3.0
times the mean noise level in the Fourier spectrum given by the middle
curve in Figure 3. We also point out that there is a low amplitude
peak at $\sim$3.85 mHz in the Fourier spectrum shown by Barlow et al.\
(2008) in their Figure 2 that could be identified with our 254.7~s
periodicity, although is not possible to infer the same for the 577.6~s
oscillation because it is embedded in the sidelobe structure of the
dominant mode in their transform (assuming that it is also present in
their data).  

\subsection{Amplitude and Phase Variations}

We investigated the stability of the amplitude and phase of the dominant
oscillation (654.4~s) and of its first harmonic (327.2~s) by performing
nightly measurements. A clue about possible amplitude variations on a
daily timescale is first provided by Figure 4, in which we show a montage
of the nightly Fourier amplitude spectra. Although the amplitude of the
first harmonic (3.0560 mHz; 327.218~s) does not change significantly as
can be judged by the eye, that of the main mode (1.5281 mHz; 654.397~s)
clearly varies in relation to the harmonic. For instance, on September 5,
the amplitude of the main mode and that of its first harmonic are about
equal, while the main mode dominates for the two other nights. Interestingly, 
similar variations in the amplitude of the main mode are also suggested
in the results of Barlow et al.\ (2008; see their Table 2). In Figure 4,
we further indicate the locations of the two other potential 
periodicities at 577.6~s and 254.7~s, but their amplitudes are so low
that possible variations are lost in the noise.

\begin{figure}[!ht]
\plotone{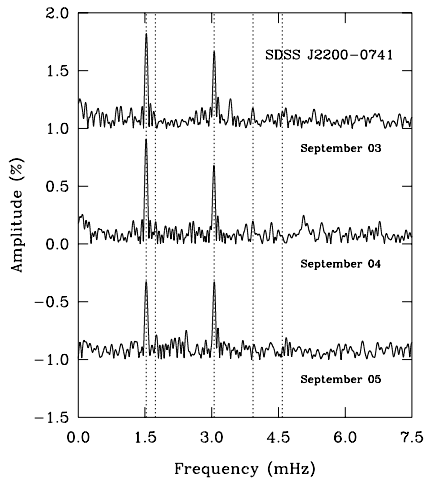}
\caption{Montage of the nightly Fourier transforms, shifted
arbitrarily along the y-axis for visualization purposes. The locations
of the five frequencies we retained are indicated by the vertical dotted
lines.}
\end{figure}

A more quantitative and standard way of measuring the nightly amplitudes
and phases is to fix the periods at their values given in Table 2 and
simultaneously perform least-squares sine fits with these periods for each
nightly run. The outputs are nightly amplitudes and phases with formal
estimates of their uncertainties. It is interesting to point out
that the formal estimates of the uncertainties on the amplitudes and
phases that came out of our least-squares exercise were, again,
essentially the same as those obtained through the method of Montgomery
\& O'Donoghue (1999), which we explicitly used after the fact as a
verification. 

Figure 5 summarizes our results in the case of the main periodicity
found in the light curve of SDSS J2200$-$0741. The upper panel in the figure
displays the amplitudes of the 654.397~s peak along with their formal 
1~$\sigma$ uncertainties for the three nightly runs. The central dotted 
horizontal line represents the weighted average of the nightly
amplitudes and the horizontal lines above and below the average value
give the 1~$\sigma$ uncertainty on that value. Likewise, the lower panel
illustrates the behavior of the phase. In the latter case, the average value
was shifted to zero (since the phase is arbitrary) and ultimately
expressed in units of cycle. The equivalent results for the 327.218~s
peak are displayed using the same format in Figure 6. As expected, the
uncertainties on the amplitudes for a given night are essentially
independent of the amplitudes themselves, while the uncertainties on the
phases increase with decreasing amplitudes. This is clearly demonstrated
by comparing the two figures.

\begin{figure}[!ht]
\plotone{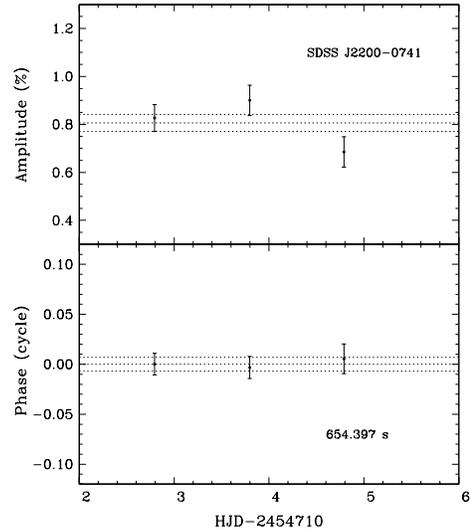}
\caption{Nightly measurements of the amplitude and phase of
the 654.397~s harmonic oscillation seen in SDSS J2200$-$0741.}
\end{figure}

\begin{figure}[!ht]
\plotone{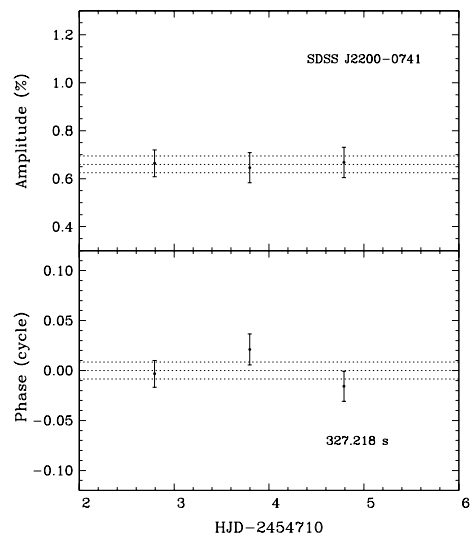}
\caption{Similar to Fig.\ 5, but for the 327.218~s
periodicity. }
\end{figure}

Quite unfortunately, no firm conclusions can be drawn from Figures 5 and
6 about possible amplitude and phase modulations in the light curve of
SDSS J2200$-$0741 on timescales of days, if we take the quantitative
results at face value. There is, however, a strong suggestion of, at
least, amplitude variation of the main mode, in accordance with the
qualitative picture that can be obtained from the nightly Fourier
transforms as depicted in Figure 4. These variations in amplitude are
probably real, but they need to be confirmed with higher sensitivity
measurements. Comparing Figure 5 with Figure 6, there is also the weaker
(but interesting) suggestion that the amplitude and phase behaviors of
the dominant periodicity may be anticorrelated with those of its first
harmonic. Clearly, however, a longer observational campaign on SDSS
J2200$-$0741 is badly needed to progress on that front.

\subsection{Pulse Shape}

Following the remarks of Montgomery et al.\ (2008) on the pulse shape in
the light curve of SDSS J1426+5752 and similar observations made by
Barlow et al.\ (2008) about SDSS J2200$-$0741 and SDSS J2348$-$0943, we
found it instructive to further investigate the question with the help
of our data set. Hence, we show in the top panel of Figure 7 our 13.29~h
long light curve of SDSS J2200$-$0741 folded on the main period of
654.397~s. To reach a decent S/N, we distributed the folded amplitudes
in 10 different phase bins, each containing 74 points on average. The
error bars about each point in the folded light curve correspond to the
errors of the mean in each bin. Given that the first harmonic of the
654.397~s mode in the light curve has a very high amplitude, about 82\%
of that of the main peak (see Table 2), it is not surprising that the
pulse shape illustrated in the top panel of Figure 7 is highly nonlinear. 
This behavior is very similar to that observed in SDSS J1426+5752, 
as can be seen in Figure 7 of Green et al.\ (2009), but is even more 
extreme here due to the higher relative value of the amplitude of the 
first harmonic.  In neither case does the folded light pulse resemble
those of typical large amplitude pulsating white dwarfs (and see Montgomery
et al.\ 2008 and Green et al.\ 2009 for specific examples).   

\begin{figure}[!ht]
\plotone{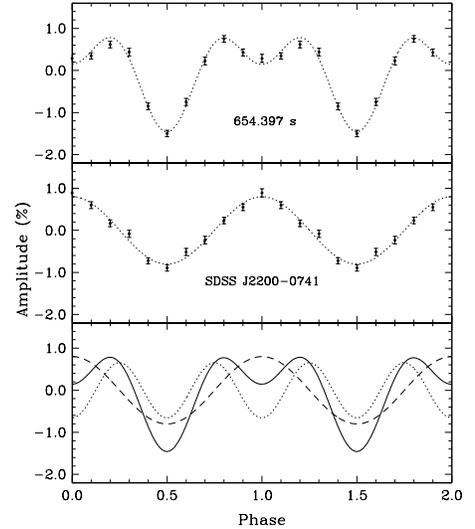}
\caption{{\it Top panel:} Light curve of SDSS J2200$-$0741 folded
on the period of 654.397~s and distributed in 10 phase bins (points with
error bars), each of which contains 74 points, on average. The dotted
curve is a model pulse shape obtained by summing two sinusoids with the
periods, amplitudes, and phases of the 654.397~s periodicity and of its
first harmonic (327.218~s) as listed in Table 2. {\it Middle panel:}
Light curve of SDSS J2200$-$0741 folded on the period of 654.397~s and
distributed in 10 phase bins after prewhitening of the first harmonic
component. The dotted curve is a pure sine wave computed using the
period, amplitude, and phase of the dominant 654.396~s periodicity. {\it
  Bottom panel:} The dotted curve is a pure sinusoid computed with the
period, amplitude, and phase of the 327.218~s harmonic; the dashed curve
is a pure sine wave computed using the period, amplitude, and phase of
the dominant 654.397~s periodicity (this is the same as the dotted curve
in the middle panel); the solid curve is the sum of these two sinusoids
and corresponds to the model pulse shape represented by the dotted curve
in the top panel.}
\end{figure}

In the middle panel of Figure 7, we again display our folded light
curve of SDSS J2200$-$0741, but only after having prewhitened the data of
the first harmonic (327.218~s) of the main periodicity. If higher order
harmonics have negligible amplitudes, and if other modes do not
interfere in the folding process, the pulse shape in the middle panel
should be that of a perfect sinusoid with an amplitude equal to that of
the 654.397~s oscillation described in Table 2. Contrary to what Green
et al.\ (2009) found for SDSS J1426+5752 (see their Figure 7), the match
between the observed points and the 654.397~s sinusoid (dotted curve) is
less than perfect here. The simplest interpretation is that higher order
terms have nonnegligible effects on the folded pulse shape, and thus we
actively sought for evidence of the presence of the second harmonic of
the main mode as described below.

In the lower panel, we plotted the same template (dashed curve)
corresponding to the 654.397~s sinusoid. Properly taking into account
the phase difference, we also plotted a sinusoid (dotted curve) with the
defining characteristics of the 327.218~s periodicity as given in Table
2. The sum of these two sine waves gives the solid curve, the overall
nonlinear pulse shape associated with the 654.397~s oscillation. We
replotted this model pulse shape in the upper panel of Figure 7 (now as a
dotted curve) so that a direct comparison can be made with the
observations. Again, the agreement is less than perfect and could be
improved as some of the points fall off the template.

To follow up on this idea of improving the model pulse shape, we went
back to the Fourier transform and searched for the possible presence of
the second harmonic of the main pulsation. This oscillation would
necessarily have a lower amplitude than 3.0 times the mean noise level
since we did not pick it up in our initial frequency analysis. In this
context, Figure 8 shows the Fourier amplitude spectrum of our light
curve of SDSS J2200$-$0741 in the immediate vicinity of where the second
harmonic of the 654.397~s periodicity should be, if present with some
amplitude. The two vertical lines define the $\pm$3~$\sigma$ frequency
range in which the second harmonic should occur. Specifically, this range
is defined by 3$\times$1.52813 $\pm$ 3$\times$(3$\times$0.00013) mHz
(see Table 2). As can be seen in the figure, there $is$ a peak with an
amplitude slightly less than 0.1\% that falls in this narrow frequency
interval. (See also the dotted vertical line segment in Figure 3
which indicates the position of this peak on another scale.) It has a
measured frequency (period) of 4.5851 mHz (218.097~s). As before, we
least-squares fitted a sine wave with that frequency to our
observational data and obtained the parameters given in Table 2.  
A statistical analysis would no doubt reveal that the
chance occurrence of a peak with an amplitude of $\sim$0.1\% (2.1 times the
mean noise level) falling within 2.3 $\mu$Hz (the formal resolution is 5.3
$\mu$Hz) of its expected position is quite small. However, in view of the 
results that we obtained, we felt that such an analysis would be redundant.
 
\begin{figure}[!ht]
\plotone{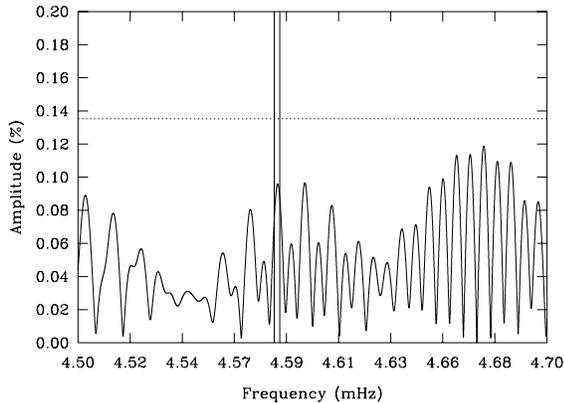}
\caption{Fourier amplitude spectrum of the light curve of 
SDSS J2200$-$0741 in the near vicinity of the second harmonic of the
main 654.397~s oscillation. The two vertical lines define the
$\pm$3~$\sigma$ frequency range in which the second harmonic should
fall. The near coincidence of a substantial frequency peak in that range
indicates that the second harmonic has been detected, albeit at an amplitude
significantly smaller than the standard detection criterion. The
horizontal dotted line defines the standard threshold of 3 times the
mean noise level (0.045\%) in the 0$-$7.5 mHz bandpass.}
\end{figure}

Indeed, Figure 9 is a remake of Figure 7, except that, this time, the
effects of the newly found 2.1~$\sigma$ peak --- interpreted as the true
second harmonic --- are taken into account. In the middle panel, the
data points correspond to the folded light curve after prewhitening the
contributions of the first (327.218~s) and second (218.097~s) harmonic
of the main mode. This time, the data points fall much closer to the
654.397~s sinusoid template. Likewise, our new model pulse shape (dotted
curve in the upper panel of Figure 9) is significantly improved at the
qualitative level compared to our initial attempt displayed in the upper
panel of Figure 7. In addition, it is revealing to point out, from the
lower panel of Figure 9, that at phase 1.0, the first harmonic is in
antiphase with the main 654.397~s periodicity, while the second harmonic
is in phase. This argues strongly against the notion that the 218.097~s
peak seen in Figure 8 could be there by chance since its phase would
presumably be a random value in that case. Together with the improvements
in the model pulse shape, we take this phase correlation as convincing 
proofs that the second harmonic of the main periodicity has indeed been
detected in our data set. It remains to be seen how the pulse shape of the
main periodicity detected in the light curve of SDSS J2200$-$0741 can be
explained and modeled in details. Along with the phase/antiphase
behavior of its harmonic components, the amplitude hierarchy of
1.00:0.82:0.12 poses an interesting challenge to modelers. 

\begin{figure}[!ht]
\plotone{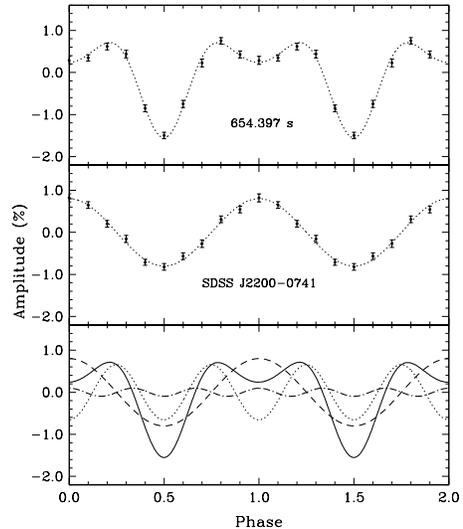}
\caption{Similar to Fig.\ 7, except that, this time, $both$
the first harmonic (dotted curve in the bottom panel) $and$ the second
harmonic (dot-dashed curve in the bottom panel) of that periodicity are
now involved in the plot. The folded light curve in the middle panel
has been obtained after prewhitening of the first and second harmonic
contributions. The model pulse shape (dotted curve in upper panel)
reproduces the observations significantly better at the qualitative
level than that shown in Fig.\ 7.}
\end{figure}

\section{SDSS J2348$-$0943}

\subsection{Observations and Frequency Analysis}

We also carried out follow-up photometric observations of SDSS
J2348$-$0943, the second variable star reported by Barlow et al.\ (2008),
and the third known pulsating Hot DQ white dwarf. SDSS J2348$-$0943 is
significantly fainter than SDSS J2200$-$0741 ($g$ = 19.00 versus $g$ =
17.70), and is not well located in the sky for observations taken at
Mount Bigelow, since it transits in the bright Tucson sky to the
south. Anticipating the need to build up sufficiently the S/N, we
requested and were allocated 17 nights of observing time in the fall
trimester of 2008. We were able to gather useful data during ten of those
nights. Table 3 gives the details. Altogether, we accumulated some 50.19~h
of observations over a moderately long timebase (1222~h, from October
1 through November 21). This corresponds to a low duty cycle of 4.1\%,
but to a relatively high temporal resolution of 0.23 $\mu$Hz. As before,
all images were taken through a broadband Schott 8612 filter. An
effective exposure time of about 69~s (on average) was chosen, and the
large field of view of the Mont4K allowed us to use nine comparison
stars for reduction purposes. This proved important for such a faint
target embedded in a relatively bright background.

% Table 3
\begin{deluxetable}{cccc}
\tablewidth{0pt}
\tablecaption{Journal of Observations for SDSS J2348$-$0943}
\tablehead{
\colhead{Date} &
\colhead{Start of Run} &
\colhead{Number of Frames} &
\colhead{Length}\\
\colhead{(UT)} &
\colhead{(HJD2454740+)} &
\colhead{} &
\colhead{(s)}
}
\startdata
2008 Oct 01 & 0.7892687 & 146 & 10,046\\
2008 Oct 02 & 1.6495160 & 322 & 22,157\\
2008 Oct 27 & 26.6075224 & 262 & 18,028\\
2008 Oct 28 & 27.5955684 & 294 & 20,230\\
2008 Oct 29 & 28.5830507 & 300 & 20,643\\
2008 Oct 30 & 29.5827912 & 296 & 20,368\\
2008 Oct 31 & 30.5891725 & 286 & 19,680\\
2008 Nov 01 & 31.5843403 & 288 & 19,817\\
2008 Nov 02 & 32.5902712 & 260 & 17,891\\
2008 Nov 21 & 51.5683732 & 172 & 11,835\\
\enddata
\end{deluxetable}

Figure 10 is a montage showing the ten nightly light curves that were
gathered. Naturally, the noise is substantially larger than what we
obtained for the brighter SDSS J2200$-$0741 (see Figure 1), but the
variability of SDSS J2348$-$0943 is nevertheless obvious in the plot. A
zoomed-in view of the light curve gathered on October 28 (2008) is
provided in Figure 11. In that case, the noise masks the pulse shape of
any dominant periodicity. Following our standard approach, we were able
to extract a  significant pulsation and a probable one in the light
curve of SDSS J2348$-$0943. Their characteristics are summarized in
Table 4 where, this time, the phase refers to the beginning of the first
run on UT 2008 October 1. The first of these pulsations, the dominant
one in amplitude, has a period of 1044.168$\pm$0.012~s, and clearly
corresponds to the single $\sim$1052~s oscillation reported by Barlow et
al.\ (2008) in their paper. Our derived amplitude for that mode,
0.81$\pm$0.07 \%, is also compatible with theirs. In addition, due to
the higher sensitivity of our data set, we likely uncovered a new mode
with a low amplitude and with a period of 416.919 $\pm$ 0.004~s which is
not harmonically related to the first one. Interestingly, we found no
sign of a strong first harmonic component of the main periodicity,
contrary to what is seen in SDSS J1426+5752 and SDSS J2200$-$0741. 

\begin{figure}[!ht]
\plotone{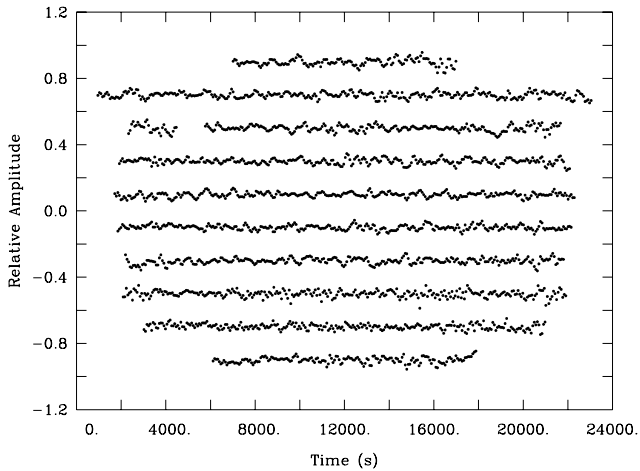}
\caption{All light curves obtained for SDSS J2348$-$0943
using the Mont4K CCD camera mounted on the Steward Observatory 1.55~m 
Kuiper telescope. The data have been shifted arbitrarily along the x and
y axes for visualization purposes. They are expressed in units of
fractional brightness intensity and seconds. From top to bottom, the
curves refer to the nights of UT October 1, October 2, October 27,
October 28, October 29, October 30, October 31, November 1, November 2,
and November 21 (2008). For details, see Table 3.}
\end{figure}

\begin{figure}[!ht]
\plotone{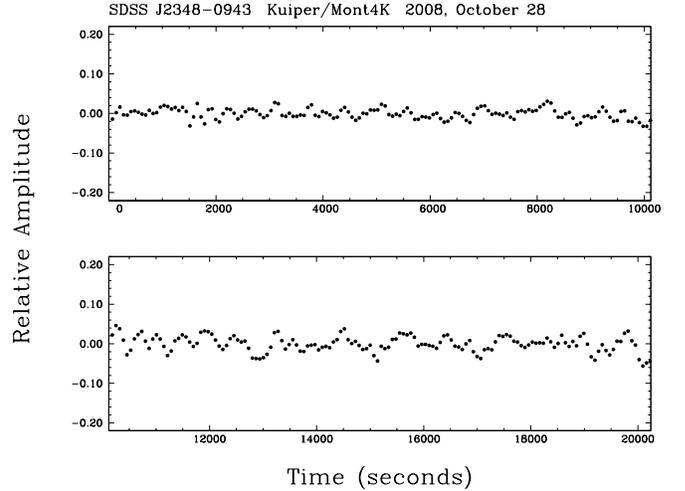}
\caption{Expanded view of the light curve of SDSS
J2348$-$0943 obtained on October 28, 2008. The units are the same as
those used in Fig.\ 10. }
\end{figure}

The upper curve in Figure 12 illustrates the Fourier amplitude spectrum
of our entire data set in the 0$-$5 mHz bandpass. The other two
transforms depict the prewhitening sequence. After removal of the two
periodicities listed in Table 4, the mean noise level found in the
residual transform (lower curve in Figure 12) corresponds to 0.088\% of
the mean brightness of the star. According to the standard criterion,
this implies that the detection of the 416.919~s periodicity is a
0.362/0.088 = 4.1~$\sigma$ result. Taking into account the number 
of available frames, we estimate that the false alarm probability for
that periodicity is $\sim$0.6\%, sufficiently small in our view for
claiming a probable detection. For its part, the 1044.168 s signal is a
9.2~$\sigma$ detection, its false alarm probability is virtually
equal to zero, and there is no doubt at all about its reality. 
Finally, we wish to point out that the estimates of the uncertainties on
the amplitudes and phases that we obtained from our least-squares
exercise in the prewhitening process turned out to be about 10\% smaller
than the estimates based on the Montgomery \& O'Donoghue (1999) method,
contrary to what we found for the brighter SDSS J2200$-$0741, for which
the two approaches gave identically the same results. We adopted the
more conservative estimates based on the Montgomery \& O'Donoghue (1999)
method in the data reported in Table 4, but we also report in
parentheses in that table the estimates based on the least-squares
approach. 

\begin{figure}[!ht]
\plotone{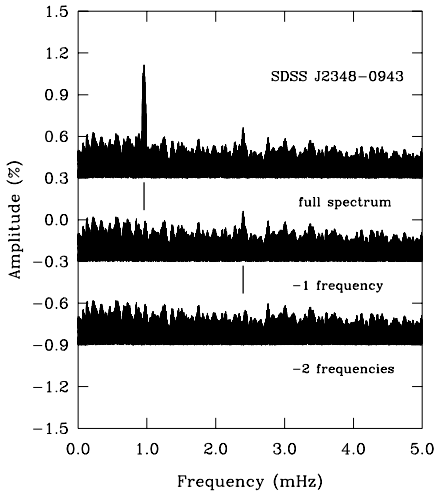}
\caption{Fourier transform of the entire data set in the
0$-$5.0 mHz range (upper curve). The lower transforms show the
successive steps of prewhitening by the strongest frequency (the
1044.168~s peak), and by the two frequencies with statistically
significant amplitudes.}
\end{figure}

\subsection{Amplitude and Phase Variations}

In a similar manner as described above for SDSS J2200$-$0741, we
investigated the possibility of detecting amplitude and phase
modulations on timescales of days by making nightly measurements. Before
presenting the results of this exercise, we draw attention to Figure 13,
which shows the Fourier transforms of four representative nights. This
plot does suggest possible amplitude variations in the two modes that we 
have isolated. However, the level of noise in each nightly transform is
rather high, which makes it difficult to be entirely convinced of
the reality of these variations. We point out, however, that the light
curve of SDSS J2348$-$0943 was completely flat as far as we could judge
in real time at the telescope during the night of November 2. And
indeed, the Fourier amplitude spectrum of that night (lower curve in
Figure 13) is entirely consistent with noise. This was rather striking
as we could always see some variations in real time during the other nights. 
And yet, there was nothing peculiar to report about the night of
November 2 in terms of weather or equipment performance.

\begin{figure}[!ht]
\plotone{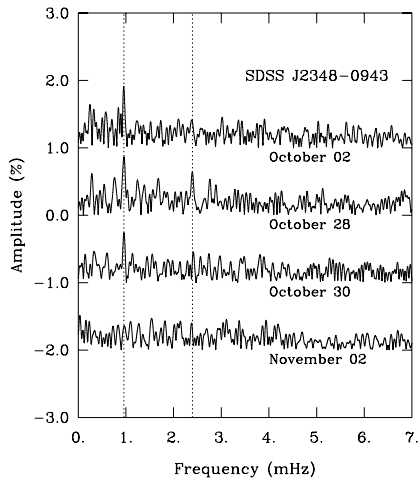}
\caption{Montage showing four of the nightly Fourier
transforms. These are shifted arbitrarily along the y-axis for
visualization purposes. The locations of the two detected frequencies
are indicated by the vertical dotted lines. Amplitude variations are
suggested.}
\end{figure}

Figure 14 summarizes our results for the daily measurements of the
amplitude and phase of the dominant 1044.168~s pulsation in the light
curve of SDSS J2348$-$0943. The format is the same as that used in
Figure 5 above.  As before, we used the Montgomery \&
O'Donoghue (1999) method for estimating the uncertainties on the nightly
values of the amplitude and phase. In the upper panel, there is a hint
of amplitude variation, particularly over the seven consecutive nights
from October 27 to November 2. As extreme cases, the amplitude of the
1044.168~s periodicity reached a maximum value of 0.939$\pm$0.147\%
during the night of October 29, and it dropped to the smallest observed
value of 0.343$\pm$0.157\% on November 2 when no obvious brightness
variations could be seen at the telescope. In comparison, within the
estimated uncertainties, the phase of the 1044.168~s oscillation appears
fairly stable. We suspect that amplitude modulations on timescales of days
probably occur for the 1044.168~s mode. Quite possibly, phase modulations 
also occur, but the accuracy with which the phase can be measured is
insufficient to be certain. 

\begin{figure}[!ht]
\plotone{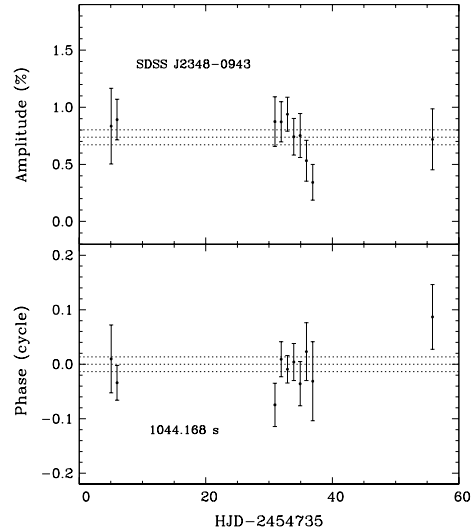}
\caption{Nightly measurements of the amplitude and phase of
the 1044.168~s harmonic oscillation seen in SDSS J2348$-$0943. }
\end{figure}

In the case of the 416.919~s oscillation, as depicted in Figure 15, the
amplitude appears not to vary significantly, but we note that the
relative errors are huge. The results for the phase are more puzzling,
but we suspect in this case that the formal uncertainties may have been
underestimated. Certainly no firm conclusions can be drawn here. As
discussed briefly below, there is interest in determining if, indeed,
the pulsations in both SDSS J2200$-$0741 and SDSS J2348$-$0943 really
display amplitude and phase variations. To reach that goal, higher
sensitivity observations will need to be gathered. In the specific case
of SDSS J2348$-$0943, a week of 4-m telescope time would probably be
needed to derive nightly measurements with sufficient accuracy.

\begin{figure}[!ht]
\plotone{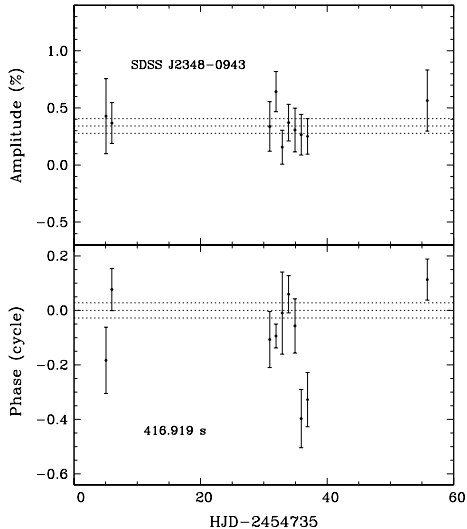}
\caption{Similar to Fig.\ 14, but for the 416.919~s
periodicity. }
\end{figure}

\subsection{Pulse Shape}

The question of the folded pulse shape in SDSS J2348$-$0943 has been
raised by Barlow et al.\ (2008), and we return to it with the help of our
higher sensitivity data set. Figure 16 shows our 50.19~h long light curve of
SDSS J2348$-$0943 folded on the main periodicity of 1044.168~s. In order
to reach an adequate S/N, we distributed the folded amplitudes in 10
distinct phase bins, each containing 263 points on average. What Figure
16 reveals is that, within our measurement errors, the pulse shape of
the 1044.168~s periodicity is perfectly linear, i.e., sinusoidal. This
is in line with the fact that no harmonics of that oscillation have been
detected in our data set. This is in marked constrast to what has been
found in the light curves of both SDSS J1426+5752 (Montgomery et
al. 2009; Green et al.\ 2009) and SDSS J2200$-$0741 (Barlow et al.\ 2008;
this work) where the first harmonic boasts an amplitude that is a very
large fraction of the amplitude of the main oscillation, thus producing
a highly nonlinear pulse shape.

\begin{figure}[!ht]
\plotone{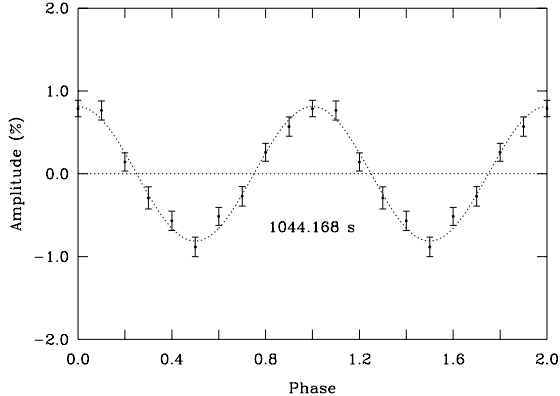}
\caption{Light curve of SDSS J2348$-$0943 folded on the
period of 1044.168~s and distributed in ten phase bins. Prewhitening of
the 416.919~s oscillation has been done prior to folding. As usual, the
curve is plotted twice for better visualization. The points give the
observational data with error bars corresponding to the errors of the
mean in each bin, which contains 263 folded points, on average. The
dotted sinusoid is a template computed on the basis of the measured
amplitude and the period of the 1044.168~s oscillation. }
\end{figure}

Given the fact that nonlinear pulse shapes have been found for the
dominant mode in both SDSS J1426+5752 and SDSS J2200$-$0741, it becomes
interesting to search for the possible presence of the first harmonic of
the main 1044.168~s periodicity, a harmonic that could perhaps be buried 
in the noise. This is similar in spirit to the procedure we followed above 
to uncover the second harmonic of the 654.397~s oscillation in SDSS
J2200$-$0741. Our search is summarized in Figure 17 which shows the
Fourier amplitude spectrum of our light curve of SDSS J2348$-$0943 in
the immediate vicinity of where the first harmonic of the 1044.168~s
periodicity should occur, if present with some amplitude. As in Figure 8,
the two vertical lines define the $\pm$3~$\sigma$ frequency range about
the first harmonic. Specifically, this range is defined 
by 2$\times$0.95770 $\pm$ 3$\times$(2$\times$0.00001) mHz (see Table 4). 
Given that the resolution (0.23 $\mu$Hz) in our SDSS J2348$-$0943 data
set is much better than that (5.30 $\mu$Hz) obtained in our shorter
time series for SDSS J2200$-$0741, the accuracy achieved in frequency is
noticeably higher, and this is reflected in the range of frequencies
plotted in Figure 17 compared to the range illustrated in Figure 8. The
frequency range ratio, 0.2:0.009, from Figure 8 to Figure 17, corresponds
approximately to the resolution ratio, 5.30:0.23, between our two data
sets. This was chosen in order to have approximately the same scale for
the band of expected harmonic frequency in both figures. 

\begin{figure}[!ht]
\plotone{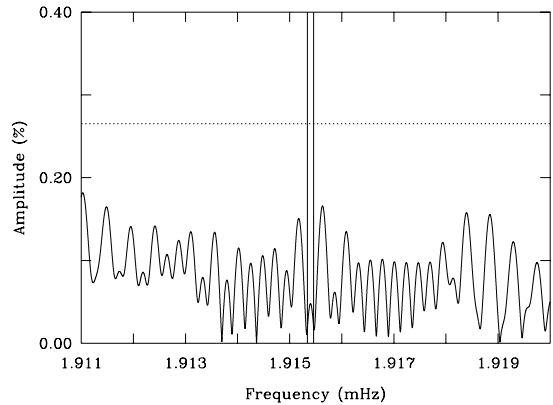}
\caption{Fourier amplitude spectrum of the light curve of 
SDSS J2348$-$0943 in the near vicinity of the first harmonic of the
main 1044.168~s oscillation. The two vertical lines define the
$\pm$3~$\sigma$ frequency range in which the harmonic should
fall. No peak in the Fourier transform with a substantial amplitude
falls in that range, thus indicating that the first harmonic of the main
periodicity is not detected in the present data set. The horizontal
dotted line defines the standard threshold of 3 times the mean noise
level (0.088\%) in the 0$-$5 mHz bandpass.}
\end{figure}

Contrary to what was found in Figure 8, Figure 17 now reveals that
there is $no$ peak with any significant amplitude in the frequency
interval where the first harmonic should be found. To be fair, there is
a tiny peak with a formal least-squares amplitude of 0.043$\pm$0.071\%. 
This corresponds to a relative amplitude of about 5\% of the amplitude
of the 1044.168~s oscillation, but the uncertainties allow a ``true'' 
amplitude that could very well be close to zero. To reinforce this result,
Figure 18 illustrates the outcome of another folding exercise in which
our light curve of SDSS J2348$-$0943, prewhitened of the two periodicities 
listed in Table 4, has been folded on the period (522.084~s) of the
potential first harmonic.  In sum, within the uncertainties, our data
show no detectable sign of an oscillation at the frequency of the first
harmonic of the main 1044.168~s mode, implying that the pulse shape of that
mode must be sinusoidal. At most, the amplitude of such a harmonic could
reach only a few percent of the amplitude of the main component, not
nearly enough to distort the pulse shape to the extent seen in SDSS
J1426+5752 and SDSS J2200$-$0741. 

\begin{figure}[!ht]
\plotone{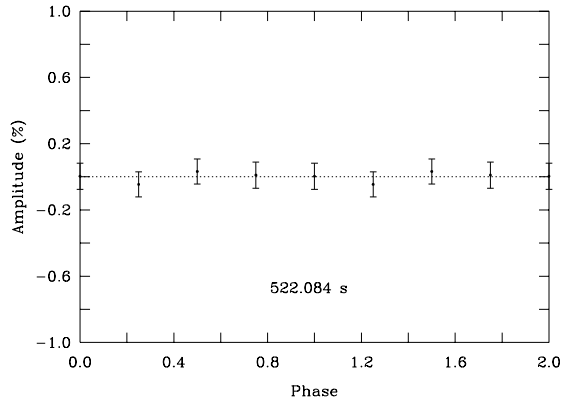}
\caption{Light curve of SDSS J2348$-$0943 folded on the
period (522.084~s) of the first harmonic of the main periodicity and
distributed in four phase bins. Within our measurement uncertainties,
no oscillation is seen at that period.}
\end{figure}

In view of this, we found the result of Barlow et al.\ (2008) on the 
pulse shape of SDSS J2348$-$0943 to be somewhat puzzling and worthy of
further investigation. Indeed, in their Figure 5, these authors
illustrate a nonlinear pulse shape for the 1044~s periodicity and use
this, among other things, to argue against the pulsational instabilities
interpretation. What is puzzling is how they could find a nonsinusoidal
pulse shape in the absence of harmonics with significant amplitudes. In
both our and their data sets, harmonics of the 1044~s oscillation have
not been detected. In such circumstances, the folded light curve has to
be an essentially perfect sinusoid, $unless$ other modes interfere in
the folding process. Such interference becomes possible, particularly
for short runs, if the contributions of the other modes are not
prewhitened before the folding process. Since Barlow et al.\ (2008)
obtained only two rather short time series on SDSS J2348$-$0943 (7446 s
on June 28 and 9520~s on July 31 2008, according to their Table 1), this
constitutes a clue on which to base a possible explanation for the small
mystery posed by their Figure 5. We risk the suggestion that the
nonsinusoidal pulse shape reported by Barlow et al.\ (2008) for the main
oscillation seen in SDSS J2348$-$0943 is due to the combination of their
short run and their neglect of the effects of the 416.919~s mode, and
not to nonlinear effects. 

\section{DISCUSSION AND CONCLUSION}

We have presented the results of follow-up photometric observations of
SDSS J2200$-$0741 and SDSS J2348$-$0943 obtained in reaction to the 
discovery of Barlow et al.\ (2008) that those two Hot DQ stars display
luminosity variations that are typical of $g$-mode oscillations in white
dwarfs. Their discovery made these stars the second and third known
specimens of a new class of pulsating white dwarfs after the GW Vir,
V777 Her, and ZZ Ceti types (see, e.g., Fontaine \& Brassard 2008), with
SDSS J1426+5752 as the prototype (Montgomery et al.\ 2008). Our
observations were gathered in integrated light using the Kuiper/Mont4K
combination at the Steward Observatory Mount Bigelow Station. By
gathering some 13.29~h of photometry on SDSS J2200$-$0741 ($g$ = 17.70)
and some 50.19~h on SDSS J2348$-$0943 ($g$ = 19.00), we were able to
reach a higher sensitivity than was possible in the discovery runs of
Barlow et al.\ (2008). 

Our results both reinforce and go beyond those of Barlow et al.\ (2008). 
We confirm that SDSS J2200$-$0741 has a light curve dominated by a
periodicity at 654.397$\pm$0.056~s along with its first harmonic
(327.218$\pm$0.017~s), in agreement with Barlow et al.\ (2008). The
amplitude of the harmonic is unusually large --- $\sim$ 82\% of the
amplitude of the main oscillation --- for a pulsating white dwarf, and
explains the highly nonlinear pulse shape of the light curve when folded
on the main period of 654.397~s. In addition, we uncovered evidence for
the presence of three new oscillations with low amplitudes, one being
the second harmonic of the main oscillation, and two others not
harmonically related to that same oscillation. Those two correspond,
presumably, to two independent pulsation modes. The first one has a
period of 577.576$\pm$0.226~s and corresponds to a 3.4~$\sigma$ result,
while the second one has a period of 254.732$\pm$0.048~s and is a
3.2~$\sigma$ detection according to  the standard criterion. Given that
the false alarm probability is not negligible for these
oscillations ($\sim$9.9\% and $\sim$27.1\%, respectively), we consider
these periodicities as possible detections only. Higher sensitivity
observations are needed to confirm their reality. The characteristics of
the five oscillations we isolated in SDSS J2200$-$0741 are summarized in
Table 2.  

We further confirm Barlow et al.'s (2008) result that the light
curve of SDSS J2348$-$0943 is dominated by an oscillation with a period
of 1044.168$\pm$0.012~s ($\sim$1052~s in their data set). However,
contrary to their findings, our light curve folded on the period of
1044.168~s shows no sign of nonlinearities. Within our measurement
errors, the pulse shape is, in fact, perfectly sinusoidal. We explicitly
searched for the possible presence of the first harmonic of the main
mode, but concluded that, if present at all, its amplitude could only
reach a few percent of the amplitude of the 1044.168~s oscillation,
quite insufficient to distort appreciably the folded light pulse. We
also uncovered evidence for the presence of a new pulsation mode in SDSS
J2348$-$0943, one with a period of 416.919$\pm$0.004~s and a low amplitude
corresponding to a 4.1~$\sigma$ detection according to the standard
criterion. The false alarm probability in this case is estimated to be
around 0.6\%, sufficiently small for us to claim a probable detection.
We suggest that the effects of that extra mode, not removed through
prewhitening in the Barlow et al.\ (2008) data, combined with the short
durations of their runs might be responsible for the nonsinusoidal 1052
s pulse shape that they discuss in their paper. The characteristics
of the two oscillations we isolated in SDSS J2348$-$0943 are summarized
in Table 4.  

% Table 4
\begin{deluxetable*}{cccc}
\tablewidth{0pt}
\tablecaption{Harmonic Oscillations Detected in the Light Curve of SDSS
  J2348$-$0943 }
\tablehead{
\colhead{Period} &
\colhead{Frequency} &
\colhead{Amplitude} &
\colhead{Phase}\\
\colhead{(s)} &
\colhead{(mHz)} &
\colhead{(\%)} &
\colhead{(s)}
}
\startdata
1044.168$\pm$0.012 & 0.95770$\pm$0.00001 & 0.812$\pm$0.071(0.065) & 879.3$\pm$14.5(13.0)\\
416.919$\pm$0.004 & 2.39855$\pm$0.00002 & 0.362$\pm$0.071(0.065) & 250.6$\pm$12.9(11.9)\\
\enddata
\end{deluxetable*}

We investigated the stability of the amplitudes and phases of the largest
amplitude frequency components that were found in the light curves of
SDSS J2200$-$0741 and SDSS J2348$-$0943. Unfortunately, the measurements 
of the nightly amplitudes and phases suffer from rather large uncertainties 
because of the noise level in the light curves of these faint
targets. We cannot therefore conclude with any certainty about the
reality of possible amplitude and phase modulations over timescales of
days. However, we note that for the dominant oscillation in both target
stars, 654.397~s in SDSS J2200$-$0741 and 1044.168~s in SDSS
J2348$-$0943, Figures 5 and 14 suggest, respectively, the strong
possibility of amplitude modulation.  This needs to be confirmed 
with higher sensitivity measurements.

It is instructive at this point to combine the results of Green et
al.\ (2009) on the first known pulsating Hot DQ white dwarf, SDSS
J1426+5752, with those of the present paper. Green et al.\ (2009) found
that the light curve of SDSS J1426+5752 is dominated by a periodicity at 
417.707~s along with its first harmonic, in agreement with the discovery
paper of Montgomery et al.\ (2008). The amplitude of the harmonic
component is some 30\% of that of the dominant oscillation, leading to a
highly nonlinear pulse shape for that periodicity. In addition,
again in agreement with the findings of Montgomery et al.\ (2008), Green
et al.\ (2009) found that the first harmonic is in antiphase with the
main component at light maximum leading to a flat maximum and a sharp
minimum in the pulse shape (and see their Figure 7). While this behavior
is contrary to what is seen in large amplitude pulsating white dwarfs of
well known types (whose pulse shape are quite generally characterized by
flat minima and sharp maxima), Green et al.\ (2009) argued that the
unusual pulse shape seen in SDSS J1426+5752 should not be taken as
evidence against pulsations as the source of the luminosity variations. 
And indeed, these authors pointed out that there exist isolated
pulsating stars, the large amplitude roAp stars, with pulse shapes 
reminiscent of those of SDSS J1426+5752. Green et al.\ (2009) further
pointed out that there is one thing in common between roAp pulsators and
the white dwarf SDSS J1426+5752 (and see Dufour et al.\ 2008c):
a large scale magnetic field sufficiently important in both cases to
disrupt the atmospheric layers and influence the pulsations there. Hence, 
the suggestion was made that the magnetic field could be responsible for
the unusual pulse shape (relative to known pulsating white dwarfs of the
GW Vir, V777 Her, and ZZ Ceti types, all of which being nonmagnetic)
observed in the light curve of SDSS J1426+5752.

The photometric properties of SDSS J2200$-$0741 are very similar to
those of SDSS J1426+5752. In particular, its light curve is dominated by
the contribution of a periodicity at 654.397~s along with that of its
first harmonic which has an amplitude equal to 82\% of that of the main
component. This leads to an even more extreme case of a highly nonlinear pulse
shape, compared to SDSS J1426+5752, when the light curve is folded on
the period of 654.397~s. Moreover, we found that the first harmonic 
is in antiphase with the main oscillation at phase 1.0 (see Figures 7
and 9 above), and this leads to a pulse shape similar to that seen in
SDSS J1426+5752, i.e., with a flat maximum and a sharp minimum. In
contrast, the pulse shape of the dominant oscillation in the light
curve of SDSS J2348$-$0943 appears perfectly sinusoidal, with no
detectable sign of its first harmonic. In view of the suggestion of
Green et al.\ (2009) about the magnetic field origin of the unusual pulse
shape seen in SDSS J1426+5752, this implies that that SDSS J2200$-$0741
is also probably magnetic, while SDSS J2348$-$0943 is not. This goes in
the same direction as the prediction made by Dufour et al. (2008b) on
the basis of purely spectroscopic arguments. 

As it turns out, SDSS J2200$-$0741 and SDSS J2348$-$0943 are the only
two known Hot DQ stars for which follow-up spectroscopy has not yet been 
carried out (see Dufour et al.\ 2009 for a report on the others). One of
us (P.D.) has developed an ongoing program at the MMT to gather higher
quality optical spectra than available in the SDSS archives, and it is
through this program that a large scale magnetic field of 1.2 MG was
detected in SDSS J1426+5752 (Dufour et al.\ 2008c). The large scale
field produces a distinct signature on the spectrum through Zeeman
splitting (see, e.g., Figure 12 of Green et al.\ 2009). Unfortunately,
possibly through some small cosmic conspiracy, the night of MMT time allocated
to observe SDSS J2200$-$0741 and SDSS J2348$-$0943 in 2008 September was
lost to bad weather. These are obvious targets for next season, but,
in the meantime, we present Figure 19, which compares the available
SDSS spectra of the three known pulsating Hot DQ white dwarfs. What Zeeman
splitting produces at this level of accuracy is a significant broadening
of the absorption lines (CII features here). Taking into account the
different S/N of the spectra (due to the different magnitudes of the
targets), it is clear that the spectrum of SDSS J2200$-$0741 resembles
that of the known magnetic star SDSS J1426+5752. In contrast, the
absorption lines in the spectrum of SDSS J2348$-$0943 have distinctly 
sharper cores than in the two other spectra. This comforts us in our
prediction that SDSS J2200$-$0741 is likely to be magnetic like SDSS
J1426+5752, and SDSS J2348$-$0943 is probably not. We are eager to
return to the MMT to find out for sure.

\begin{figure}[!ht]
\plotone{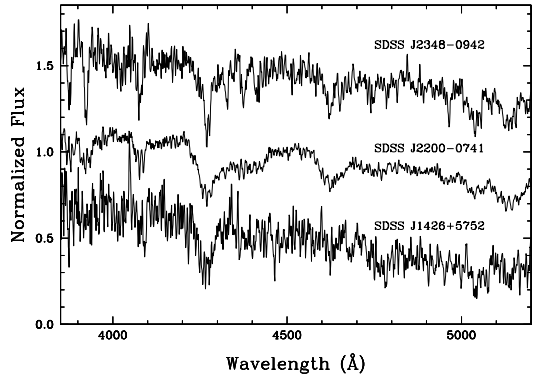}
\caption{Archived SDSS optical spectra of the three known
pulsating carbon-atmosphere white dwarfs. The absorption lines are
noticeably sharper in SDSS J2348$-$0943 than in the other two stars.
Note that, for better clarity, we applied a three-point average window
smoothing in the display of these spectroscopic data.}
\end{figure}

If indeed SDSS J2200$-$0741 turns out to be a magnetic star, it will
become, because of its relative brightness, the best test bed for the
idea that some of these stars could be white dwarf analogs of roAp
stars. In a roAp object, amplitude modulation is associated with the
rotation of the star (see, e.g., Kurtz 1990). The pulsations align
themselves along the magnetic field axis which is itself inclined with
respect to the rotation axis of the star. The viewing aspect of the
pulsations thus changes periodically with rotation, which produces
amplitude modulation of a given mode. It will therefore be of utmost
interest to verify whether rotationally-induced amplitude modulation can
be observed in SDSS J2200$-$0741, which would make it a true white dwarf
equivalent of a roAp star. 

Finally, we point out that the Hot DQ white dwarfs remain as puzzling as
ever. According to Dufour et al. (2009), at least four (and perhaps six)
of the ten known Hot DQ's are magnetic. This is much higher than the
proportion of about 10\% found in the normal population of white dwarfs
according to Liebert, Bergeron, \& Holberg (2003). And if we are correct
in the interpretation presented in this paper, there are three pulsating
Hot DQ stars out of a sample of eight investigated so far, two of which
are magnetic and the other not.

\acknowledgements

This work was supported in part by the NSERC of Canada. P. Dufour is a
CRAQ postdoctoral fellow. G. Fontaine also acknowledges the contribution
of the Canada Research Chair Program.  

% References

\end{document}